\documentclass{ajour}
\usepackage{amssymb} 
\usepackage{amsmath} 

\begin{document}

%%%%% To be entered at Academic Press: =====>>
%
% Uncomment line below only when doing final typesetting,
%\finaltypesetting
% \journame{}
% \articlenumber{}
% \yearofpublication{}
% \volume{}
% \cccline{}
% \received{}
% \revised{}
% \accepted{}

% communication line, use: \commline{Communicated by...}
% \commline{Communicated by... }

\authorrunninghead{Cari\~nena, Fern\'andez and Ramos}
\titlerunninghead{Group approach to the intertwined Hamiltonians}

%% To set particular page number:
%\setcounter{page}{261} %%

%% <<== End of commands to be entered at Academic Press

%%  Authors, start here ==>>

%\draft % Optional, will cause a line at the bottom of each page
%% with the words `Draft' and the time and date that the article
%% was LaTeXed. Will also double space text.

%%%%%%%%%%%%%%%%%%%%%%%%%%%%%%%%%%%%%%%%%%%%%%%%%%%%%%%%%
%                                                       %
%               Author's definitions                    %
%                                                       %
%%%%%%%%%%%%%%%%%%%%%%%%%%%%%%%%%%%%%%%%%%%%%%%%%%%%%%%%%

\def\ba{\begin{eqnarray}}
\def\ea{\end{eqnarray}}

\def\R{\mathbb R}
\def\C{\mathbb C}
\def\Z{\mathbb Z}
\def\N{\mathbb N}
\def\a{\alpha}                  % alpha
\def\b{\beta}                   % beta
\def\g{\gamma}                  % gamma
\def\d{\delta}                  % delta
\def\e{\epsilon}                % epsilon
\def\l{\lambda}                 % lambda
\def\p{\phi}                    % phi
\def\ps{\psi}                   % psi
\def\vp{\varphi}                % varphi
\def\z{\zeta}                   % zeta

\def\GR{{\cal G}}

\def\wt{\widetilde}
\def\Map{\mathop{\rm Map}\nolimits}

\def\matriz#1#2{ \left( \begin{array} {#1} #2 \end{array}\right)}

\newcommand{\bea}{\begin{eqnarray}}
\newcommand{\eea}{\end{eqnarray}}

\font\tengoth=eufm10 \font\sevengoth=eufm7 \font\fivegoth=eufm5
\newfam\gothfam
\textfont\gothfam=\tengoth \scriptfont\gothfam=\sevengoth
  \scriptscriptfont\gothfam=\fivegoth
  \def\goth{\fam\gothfam}               %% Euler Fraktur (math mode only)
\font\frak=eufm10 scaled\magstep1
\def\goth #1{\hbox{{\frak #1}}}

%%%%%%%%%%%%%%%%%%%%%%%%%%%%%%%%%%%%%%%%%%%%%%%%%%%%%%%%%
%                                                       %
%             End of author's definitions               %
%                                                       %
%%%%%%%%%%%%%%%%%%%%%%%%%%%%%%%%%%%%%%%%%%%%%%%%%%%%%%%%%

\title{Group theoretical approach to the intertwined Hamiltonians
%\footnote{Manuscript pages: 51. No figures. No tables.} 
%\thanks{Proposed running head: Group approach to 
%the intertwined Hamiltonians}
}

%\subtitle{}

\author{Jos\'e F. Cari\~nena
%\thanks{Corresponding author:
%see adress and e-mail on first page. Fax: +34 976 76 12 64.}
and Arturo Ramos}
\affil{Departamento de F\'{\i}sica Te\'orica. Facultad de Ciencias. \\
Universidad de Zaragoza, 50009, Zaragoza, Spain.}
\email{jfc@posta.unizar.es and arrg@wigner.unizar.es}
\and
\author{David J. Fern\'andez C.}
\affil{Departamento de F\'{\i}sica, CINVESTAV--IPN, \\A.P. 14--740,
07000 M\'exico D.F., M\'exico. }
\email{david@fis.cinvestav.mx}

\abstract{
We show that the finite difference B\"acklund formula for the 
Schr\"odinger Hamiltonians is a particular element of the 
transformation group on the set of Riccati equations considered 
by two of us in a previous paper.
Then, we give a group theoretical explanation 
to the problem of Hamiltonians related by a first order differential operator. 
A generalization of the finite difference algorithm relating 
eigenfunctions of \emph{three} different Hamiltonians is found, 
and some illustrative examples of the theory are
analyzed, finding new potentials for which one eigenfunction and its 
corresponding eigenvalue is exactly known. 
}


\begin{article}

\section{Introduction}

The factorization of Schr\"odinger Hamiltonians in terms of differential
operators of first order plays an important r\^ole in the search of
quantum systems for which the energy spectrum is completely known
\cite{InfHul51,GenKriv85,CarMarPerRan98}.  It is closely related with the
existence of intertwining operators \cite{FerHusMiel98,FerHus99},
Supersymmetric Quantum Mechanics \cite{CoopKhaSuk95} and Darboux
transformations \cite{MatSal91}, among other questions. Indeed, most of
the exactly solvable potentials can be obtained by making use of an
appropriate intertwining operator transformation. However, nowadays the
point is not just to factorize some well-known Schr\"odinger Hamiltonians
\cite{InfHul51} but to generate new solvable ones. The key to this
technique is to intertwine two different Hamiltonians by a differential
operator $A$, usually of first order 
\cite{CarMarPerRan98,FerHusMiel98,FerHus99,JunRoy98,CarRam2,CarRam3} or higher
\cite{AndIofSpi93,AndIofNis95,aicd95,
bs95,BagSam97,Fer97,FerGlaNie98,AndCanIofNis00}, 
provided one of them is solvable. Concerning iterations
of the first order intertwining technique, 
one of the authors (DJF), Hussin and
Mielnik \cite{FerHusMiel98} have recently used a \emph{finite difference
algorithm} which provides algebraically the solution of the key Riccati
equation at a given iteration step in terms of two solutions of the
corresponding Riccati equation at the previous step associated to two
different `factorization energies'. This procedure has been successfully
applied in order to obtain new exactly solvable Hamiltonians departing
from the harmonic oscillator and Coulomb potentials
\cite{FerHusMiel98,FerHus99,Ros}. 

On the other hand, a number of authors have studied the 
transformation groups related with Riccati equations
\cite{And80,HarWinAnd83,OlmRodWin8687,
BecHusWin86,BecHusWin86b,BecGagHusWin90,HavPosWin99,CarRam99}.
In the last of these references, it has been used a 
transformation group on the set of Riccati equations 
in order to analyze their integrability.
We think that this group theoretic approach could 
shed light on the abovementioned problem of intertwined 
or $A$--related Hamiltonians. A natural question is
whether there is a relation of this group action with the finite
difference algorithm. 

We will try to answer these questions in this article, which is organized
as follows. In Section 2 we recall briefly the problem of two Hamiltonians
intertwined by a \emph{first} order differential operator, a technique
which is also known as $A$--related Hamiltonians. In Section 3 we relate
the Schr\"odinger equations arising in the previous problem with certain
Riccati equations by making use of the classical Lie theory of reduction
by infinitesimal symmetries of differential equations, 
as applied to homogeneous linear
second-order equations. In Section 4 we recall the action of a group on
the set of Riccati equations described in \cite{CarRam99}. Using this
technique we will prove the finite difference theorem \cite{FerHusMiel98} 
in a way alternative to \cite{mnro00}. In Section 5 we will
determine the elements of the transformation group preserving the
subset of Riccati equations arising from the set of Schr\"odinger
equations after applying the reduction process outlined in Section 3. 
We will find a new transformation relating \emph{three} different
Schr\"odinger equations, which represents a generalization of the finite
difference B\"acklund algorithm. 
In Section 6 we find that
the problem of $A$--related Hamiltonians can be explained exactly in terms
of the transformation group on the set of Riccati equations and the
reduction procedure of Section 3. Section 7 illustrates the use
of the new Theorems of Section 5 in the search of potentials 
for which one eigenstate and the corresponding eigenvalue will be exactly known. 
In particular, Examples~\ref{ex_oscil_inter}, 
\ref{coul_variant_1} and \ref{ex_Coul_2}
will provide potentials 
essentially different from the original ones. 
Finally, we give in Section 8 some conclusions and
an outlook for future work. 

\section{Hamiltonians related by first-order differential operators.
\label{relHamiltonians}}

The simplest way of generating an exactly solvable Hamiltonian $\widetilde
H$ from a known one $H$ is just to consider an invertible bounded operator
$B$, with bounded inverse, and defining $\widetilde H=BHB^{-1}$. This
transformed Hamiltonian $\wt H$ has the same spectrum as the starting one
$H$. As a generalization (see e.g.  \cite{CarMarPerRan98}), we will say
that two Hamiltonians $H$ and $\wt H$ are intertwined or 
$A$--related when $AH=\wt HA$, where $A$ may have no inverse. 
In this case, if $\psi $ is an eigenvector of
$H$ corresponding to the eigenvalue $E$ and $A\psi\not =0$, at least
formally $A\psi$ is also an eigenvector of $\wt H$ corresponding to the
same eigenvalue $E$.

If $A$ is a first order differential operator,
\ba
A=\frac {d}{dx}+W(x)\ ,\quad\quad\mbox{and}\quad\quad 
A^{\dagger}=-\frac{d}{dx}+W(x)\ ,
\label{defAAdag}
\ea
then the relation $AH=\wt HA$, with
\begin{equation}
H=-\frac{d^2}{dx^2}+V(x)\,,\qquad \wt H=-\frac{d^2}{dx^2}+\wt V(x)\ ,
\label{defHHtil}
\end{equation}
lead to
$$V=-2W'+\wt V, \qquad W(V-\wt V)=-W''-V'\ .
$$
Taking into account the first equation, the second becomes $2WW'=W''+V'$,
which can easily be integrated giving
\begin{equation}
V=W^2-W' + \e\,,         \label{ricV} 
\end{equation}
and then,
\begin{equation}
\wt V=W^2+W' + \e\,,     \label{ricVtil}
\end{equation}
where $\e$ is an integration constant. The important point here is that
$H$ and $\wt H$, given by (\ref{defHHtil}), are related by a first order
differential operator $A$, given by (\ref{defAAdag}), if and only if there
exist a constant $\e$ and a function $W$ such that the pair of Riccati
equations (\ref{ricV}) and (\ref{ricVtil}) are satisfied
\emph{simultaneously}.  Moreover, this means that both Hamiltonians can be
factorized as
\begin{equation}
H=A^{\dag}A+\e\,,\qquad \wt H=AA^{\dag}+\e\ .\label{factorHHtil}
\end{equation}

Adding and subtracting equations (\ref{ricV}) and (\ref{ricVtil})
we obtain the equivalent pair which relates $V$ and $\wt V$
\begin{eqnarray}
\wt V-\e&=&-(V-\e)+2W^2\,,       \label{relVVtilcuad} \\
\wt V&=&V+2 W'\,.              \label{relVVtilder} 
\end{eqnarray}
The function $W$ satisfying these equations is usually called
\emph{superpotential}, the constant $\e$ is the \emph{factorization energy}
or \emph{factorization constant} and $\wt V$ and $V$ (resp. $\wt H$ and
$H$) are said to be \emph{partner} potentials (resp. Hamiltonians). 

Notice that the initial solvable Hamiltonian can indistinctly be chosen as
$H$ or $\wt H$. In both cases the point will be to find a solution $W$ of
the corresponding Riccati equation (\ref{ricV}) or (\ref{ricVtil}) for a
specific factorization energy $\e$. {}From this solution the expression for
the (possibly) new potential follows immediately from (\ref{relVVtilder}).

\section{Dilation symmetry and reduction of a linear second-order 
differential equation \label{sode_Ricc}}

Let us briefly recall a well-known method of relating a homogeneous
linear second-order differential equation to a Riccati equation, which
can be regarded as an application of the classical Lie theory of
infinitesimal symmetries of differential equations.  Its importance will
become clear when applying the method to time-independent Schr\"odinger
equations. 

The homogeneous linear second-order differential equation
\begin{equation}
\frac{d^2 z}{dx^2}+b(x)\frac{d z}{dx}+c(x)z=0\,,        \label{eq2ord}
\end{equation}
admits as an infinitesimal symmetry the vector field 
$X=z\,\partial/{\partial z}$ generating dilations 
(see e.g. \cite{CarMarNas}) in the variable $z$, which is defined for $z\neq 0$.
According to the Lie theory of infinitesimal symmetries 
of differential equations, we
should  change the coordinate $z$ to a new one, $u=\varphi(z)$, such
that the vector field $X=z\,{\partial}/{\partial z}$ becomes a translation 
generator $X={\partial}/{\partial u}$ in the new variable.
This change is determined by the
equation $Xu=1$, which leads to $u=\log |z|$, i.e. $|z|=e^{u}$. In both cases
of regions with $z>0$ or $z<0$ we have 
$$
\frac{dz}{dx}=z\frac{du}{dx}\,,\quad\quad\mbox{and}
\quad\quad\frac{d^2 z}{dx^2}=z\bigg(\frac{du}{dx}\bigg)^2+z\,\frac{d^2 u}{dx^2}\,,
$$
so the equation (\ref{eq2ord}) becomes
\begin{equation}
\frac{d^2 u}{dx^2}+b(x)\frac{d u}{dx}+\left(\frac{du}{dx}\right)^2+c(x)=0\,.
\nonumber
\end{equation}
As the unknown function $u$ does not appear in the preceding equation but
just its derivative, we can lower the order by introducing the new 
variable $w={du}/{dx}$. We arrive to the following Riccati equation for $w$ 
\begin{equation}
\frac{dw}{dx}=-w^2-b(x)w-c(x)\,.
\label{ric1}
\end{equation}
Notice that from $dz/dx=z\,du/dx$ and the definition of $w$ we have 
\begin{equation}
w=\frac 1 z\, \frac{dz}{dx}\,.                  \label{cam1}
\end{equation}
The second order differential equation (\ref{eq2ord}) is equivalent 
to the set of (\ref{ric1}) and (\ref{cam1}), because
given a function $w$ satisfying 
(\ref{ric1}), the function $z$ defined (up to a factor) by (\ref{cam1}),
i.e. $z(x)=\exp\left(\int^x w(\zeta)\, d\zeta\right)$,
satisfies (\ref{eq2ord}). We could have followed a similar 
pattern straightening out the vector 
field in the opposite sense, that is, by imposing $Xu=-1$.
This would have lead to $u=-\log |z|$, or $|z|=e^{-u}$. 
Now, in either case
of $z>0$ or $z<0$ we have ${dz}/{dx}=-z\,{du}/{dx}$ and 
${d^2 z}/{dx^2}=z\,({du}/{dx})^2-z\,{d^2 u}/{dx^2}$,
so we finally obtain the Riccati equation
\ba
\frac{dw}{dx}=w^2-b(x)w+c(x)\,,                 \label{ric2}
\ea 
where now 
\ba
w=\frac{du}{dx}=-\frac 1 z\, \frac{dz}{dx}\,.
\label{cam2}
\ea

We will distinguish in what follows
between these two alternatives of reduction
of (\ref{eq2ord}) by means of a subscript 
$+$ or $-$ in the corresponding functions, respectively. 
We remark that both are defined \emph{locally}, that is, 
in open intervals where $z$ has a constant sign.

Let us apply these ideas to the particular case of the one-dimensional
time-independent Schr\"odinger equation
\begin{equation}
-\frac{d^2 \phi}{dx^2}+(V(x)-\e)\phi=0\,,       \label{Schr}
\end{equation}
where $V(x)$ is the potential and $\e$ is some specific energy eigenvalue.
As explained before, we can reduce (\ref{Schr}) either to the pair
\begin{equation}
W_+^\prime=-W_+^2+(V(x)-\e)\,, \quad W_+=\frac 1 \phi\frac{d\phi}{dx}\,,
                                                \label{pair1_Schr}
\end{equation}
or, alternatively, to the pair
\begin{equation}
W_-^\prime=W_-^2-(V(x)-\e)\,,  \quad W_-=-\frac 1 \phi\,\frac{d\phi}{dx}\,. 
                                                \label{pair2_Schr}
\end{equation}
The Riccati equations appearing in these pairs resemble those appearing in
Section 1, namely equations (\ref{ricV}) and (\ref{ricVtil}), but in the
systems (\ref{pair1_Schr}) and (\ref{pair2_Schr}) the unknown functions
$W_+$ and $W_-$ are related by $W_+=-W_-$, while in both (\ref{ricV}) and
(\ref{ricVtil}) the unknown $W$ is \emph{the same} function.

However, the previous remark will be useful in the interpretation of
equations (\ref{ricV}) and (\ref{ricVtil}). We can rewrite them as
\ba
W^\prime&=&W^2-(V(x)-\e)\,,      \label{RicV}\\
W^\prime&=&-W^2+(\wt V(x)-\e)\,.        \label{RicVtil}
\ea 
Then, we can regard equation (\ref{RicV}) (resp. equation (\ref{RicVtil})) 
as coming from a Schr\"odinger-type equation like (\ref{Schr}) (resp.
like $-\,{d^2 \wt \phi}/{dx^2}+(\wt V(x)-\e)\wt\phi=0$) by means of,
respectively, the changes
\begin{equation}
W=-\frac 1 \phi\,\frac{d\phi}{dx}\,,\quad\quad\mbox{or} 
\quad\quad
W=\frac 1 {\wt\phi}\,\frac{d\wt\phi}{dx}\,,     \label{chW1}
\end{equation}
so the two \lq\lq eigenfunctions\rq\rq\ $\phi$ and $\wt\phi$ of the
mentioned Schr\"odinger-type equations are related by
$\phi\wt\phi=\mbox{Const.}$ Of course, the changes (\ref{chW1}) are
defined locally, i.e. in common open intervals of the domains of $\phi$
and $\wt\phi$ determined by two consecutive zeros of $\phi$ or $\wt\phi$, or
maybe by a zero and a boundary of the domain of the problem.  Note
that there is no reason why they should provide functions $\phi$,
$\wt\phi$ defined in the same way in the entire domain of $W$, but in
general they will be defined interval-wise. Moreover, if we choose the
function $W$ of the two operators $A$ and $A^\dagger$ defined in
(\ref{defAAdag}) as given by (\ref{chW1}) it holds $A\phi=0$ and
$A^\dagger\wt\phi=0$.

We have seen that the Riccati equations (\ref{RicV}) and (\ref{RicVtil}) 
correspond by means of the changes (\ref{chW1})  to two
Schr\"odinger-type equations which in turn are equivalent to
\begin{equation}
H\phi=\e\,\phi\,,\ \ \ \wt H\wt\phi=\e\,\wt\phi\,, 
\label{Hequats}
\end{equation}
where $H$ and $\wt H$ are given by (\ref{defHHtil}).  Then, it is
equivalent to say that $H$ and $\wt H$ are $A$--related, with associated
constant $\e$, to say that the pair of functions $\phi$ and $\wt\phi$,
which satisfy $\phi\wt\phi=\mbox{Const.}$, are respective eigenfunctions
with eigenvalue $\e$ of the Hamiltonians $H$ and $\wt H$.  Each of these
facts imply that both Hamiltonians can be factorized as in
(\ref{factorHHtil}).  We finally remark again that these factorizations
make sense only locally, i.e. in common open intervals where $\phi$ and
$\wt\phi$ are defined. 

A special case where all becomes globally defined arises when $\phi$ or
$\wt\phi$ is the ground state wave function of its respective Hamiltonian,
having then no zeros in the entire domain of the problem. Then, remembering
the equation (\ref{relVVtilder}) it is clear the relation of what we have
just exposed with the Darboux transformations in the context of 
one-dimensional or Supersymmetric Quantum Mechanics, as it is shown, for example,
in \cite[pp. 7, 24]{MatSal91}.

Moreover, for $\e$ below the ground state energy of $H$ (resp. $\wt H$) 
it is sometimes possible to find a non-normalizable eigenfunction
$\phi$ (resp. $\wt\phi$) of $H$ (resp. $\wt H$) without zeros, 
leading to physically interesting potentials \cite{FerHusMiel98,FerHus99}.
However, as this procedure is slightly involved, we shall restrict for simplicity
the discussions along this article to the previous case of $\e$ equal  
to the ground state energy of $H$ or $\wt H$.

\section{Transformation group on the set of Riccati 
equations\label{trgrsetRicceqs}}  

In this section we introduce the technique of relating different Riccati
equations by means of a group action. The procedure, of key importance in
what follows, was described in \cite{CarRam99} in order to give a group
theoretical explanation of some properties of the Riccati equation.

We recall that a general Riccati equation is the non-linear 
first order differential equation
\begin{equation}
\frac{dy(x)}{dx}=a_2(x)\,y^2(x)+a_1(x)\,y(x)+a_0(x)\,,
\label{Riceq}
\end{equation}
where the coefficients $a_i(x)$, $i=0,\,1,\,2$, are smooth functions of the
independent variable $x$. A particular Riccati equation, determined by a
specific election of these coefficient functions, can be regarded as a
curve in ${\R}^3$, i.e., an element of $\Map({\R},\,{\R}^3)$.

We can transform every function $y(x)$ in the extended real line
$\overline\R$ by an element of the group of smooth $SL(2, {\R})$--valued
curves $\Map({\R},\,SL(2,{\R}))$, which from now on will be denoted
${\GR}$, by means of the action
$\Theta:\GR\times\Map({\R},\,{\overline\R})\rightarrow\Map({\R},\,{\overline\R}
)$ defined as follows \cite{CarRam99}:
\bea
\Theta(A(x),y(x))={\frac{\a(x) y(x)+\b(x)}{\g(x) y(x)+\d(x)}}\ ,\ \ \
\mbox{if\ }\ y(x)\neq-{\frac{\d(x)}{\g(x)}}\ ,
\label{actTheta1}\\
\Theta(A(x),\infty)={\frac{\a(x)}{\g(x)}}\ ,\ \ \ \ 
\Theta\bigg(A(x),-{\frac{\d(x)}{\g(x)}}\bigg)=\infty\ ,
\label{actTheta2}
\eea
when 
\bea
A(x)=\matriz{cc} {{\a(x)}&{\b(x)}\\{\g(x)}&{\d(x)}}\,\in{\cal G}\ .
\label{Agauge}
\eea
It is easy to check that the Riccati equation (\ref{Riceq})
transforms under the generic transformation $\overline y(x)=\Theta(A(x),y(x))$
into a new  Riccati equation for $\overline y(x)$
with coefficients $\overline a_i(x)$, $i=0,\,1,\,2$. The relation amongst the 
new and the old coefficients is given by
\bea
\overline a_2&=&{\d}^2\,a_2-\d\g\,a_1+{\g}^2\,a_0+\g {{\d}^\prime}- \d
{\g}^\prime\,, \label{ta2}\\
\overline a_1&=&-2\,\b\d\,a_2+(\a\d+\b\g)\,a_1-2\,\a\g\,a_0   
+ \d {\a}^\prime-\a {\d}^\prime+\b {\g}^\prime-\g {\b}^\prime\,,
\label{ta1}\\
\overline a_0&=&{\b}^2\,a_2-\a\b\,a_1+{\a}^2\,a_0+\a{\b}^\prime-
\b{\a}^\prime  \,,
\label{ta0} 
\eea
where the prime denotes derivative respect to $x$.  We would like to
remark that similar transformations, using $GL(2,{\R})$ instead of
$SL(2,{\R})$ have been used in \cite{Cal63,Str91}, also in connection with
transformations of the Riccati equation. However, it seems that none of
them noticed that these transformations constitute indeed a group action,
as we will see next. 

The previous relation can be written in a matrix form
\bea
\matriz{c}{\overline a_2\\ \overline a_1\\ \overline a_0} 
&=&\matriz{ccc}
{{\d}^2&{-\d\g}&{{\g}^2} \\ {-2\,
\b\d}&{\a\d+\b\g}&{-2\,\a\g}\\{{\b}^2}&{-\a\b}&{{\a}^2}}
\matriz{c}{{a_2}\\
{a_1}\\{a_0}}
\nonumber\\     
& &+\matriz{c}{\g {{\d}^\prime}-\d {\g}^\prime\\
\d {\a}^\prime-\a {\d}^\prime+\b {\g}^\prime-\g {\b}^\prime\\
 \a {\b}^\prime-\b {\a}^\prime}\ .
\label{transf_ai_dept_matricial}
\eea
If we define the maps
\ba
B(A)&=&\matriz{ccc}
{{\d}^2&{-\d\g}&{{\g}^2} \\ {-2\,
\b\d}&{\a\d+\b\g}&{-2\,\a\g}\\{{\b}^2}&{-\a\b}&{{\a}^2}}\ , \\
\theta(A)&=&\matriz{c}{\g {\d}^\prime-\d {\g}^\prime\\
\d {\a}^\prime-\a {\d}^\prime+\b {\g}^\prime-\g {\b}^\prime\\
                \a {\b}^\prime-\b {\a}^\prime }\ , 
\ea
where $A\in\GR$, it can be easily checked that $B$ is a linear
representation of the group $\GR$ on $\Map({\R},\,{\R} ^3)$.  The
restriction of $B$ to the subgroup of constant $SL(2,\,\R)$--valued curves
is nothing but the adjoint action of $SL(2,\,\R)$ on its Lie algebra
\cite{CarRam99}.  On the other hand, it can be proved that $\theta$
defines an 1--cocycle for $B$, which means that (see \cite{CarRam99,LM87})
\begin{equation}
\theta(A_1 A_2)=B(A_1)(\theta(A_2))+\theta(A_1)\,,
\quad\quad \forall\,A_1,\,A_2\in\GR\,,
\end{equation}
and moreover $\theta$ is {\sl not\/} an 1--coboundary for $B$. 
Consequently, see e.g \cite{LM87}, the expression
(\ref{transf_ai_dept_matricial}) defines an affine action of ${\GR}$ on
the set of general Riccati equations, which in turn can be identified with
the set of curves on the Lie algebra ${\goth {sl}}(2,\,\R)$. 

In practical terms, all this means that the composition of two
transformations of type (\ref{transf_ai_dept_matricial})  on the set of
Riccati equations is exactly the same as making only one transformation in
the following way. If $T_A$ denotes the transformation of type
(\ref{transf_ai_dept_matricial}) associated with $A\in\GR$, then it holds
\begin{equation}
T_{A_1}\circ T_{A_2}=T_{A_1 A_2}\,,\quad\quad\forall\,A_1,\,A_2\in\GR\,.
\end{equation}
It is possible to generalize this affine action in more general
situations, when an arbitrary finite dimensional Lie group is involved
\cite{CarGraRam}. 

In short, if $y(x)$ is a solution of the equation (\ref{Riceq}) and 
we transform it by means of $\overline y(x)=\Theta(A(x),y(x))$, then 
$\overline y(x)$ will be a solution of the Riccati equation with
coefficient functions (\ref{ta2}), (\ref{ta1}) and (\ref{ta0}).
By means of this technique we will prove the following 
result (the notation has been adapted to the one used along this article). 

\begin{theorem}[Finite difference B\"acklund algorithm  
\cite{FerHusMiel98,mnro00,Adl93,Adl94}] 
\par
Let $w_k(x)$, $w_l(x)$ be two solutions of the Riccati equations
$w^\prime+w^2=V(x)-\e_k$ and $w^\prime+w^2=V(x)-\e_l$, respectively,
where $\e_k< \e_l$. Then the function $w_{kl}(x)$ defined by
\ba
w_{kl}(x)=-w_k(x)-\frac{\e_k-\e_l}{w_k(x)-w_l(x)}\,,
\label{betakl}
\ea
is a solution of the Riccati equation
$w^\prime+w^2=V(x)-2\,w_k^\prime(x)-\e_l$. 
\label{FerHusMiel_theorem}
\end{theorem}

\begin{proof}
The function $w_l(x)$ satisfies the Riccati equation
$w^\prime+w^2=V(x)-\e_l$ by hypothesis. We transform it by means of the
element $A_0(x)$ of $\GR$ given by
\begin{equation}
A_0(x)=\frac{1}{\sqrt{a}}
\matriz{cc}{{h(x)}&{-h^2(x)+a}\\{-1}&{h(x)}}\,,
\label{transA0}
\end{equation}
where $h(x)$ and $a$ are a function with the same domain as $w_l(x)$ 
and a \emph{positive} constant, respectively.
Notice that $A_0(x)\in\GR$ since its determinant is always one, for all $x$
in the domain of $h(x)$. According to (\ref{actTheta1}) and
(\ref{actTheta2}) we compute
\begin{equation}
\Theta(A_0(x),w_l(x))=\frac{h(x)w_l(x)-h^2(x)+a}{h(x)-w_l(x)}
=-h(x)+\frac{a}{h(x)-w_l(x)}\,.  \nonumber
\end{equation}
This is a solution of the Riccati equation with coefficient functions
given by (\ref{ta2}), (\ref{ta1}) and (\ref{ta0}), with
matrix elements $\a(x)=\d(x)=h(x)/\sqrt{a}$, $\b(x)=(-h^2(x)+a)/\sqrt{a}$,
$\g(x)=-1/\sqrt{a}$ and coefficients of the initial Riccati equation $a_2(x)=-1$,
$a_1(x)=0$ and $a_0(x)=V(x)-\e_l$.
Simply performing the operations, we find
\ba
\overline a_2(x)&=&\frac{1}{a}\{-h^2(x)-h^\prime(x)+ 
V(x)-\e_l+a\}-1\,,                   \nonumber\\
\overline
a_1(x)&=&\frac{2\,h(x)}{a}\{-h^2(x)-h^\prime(x)+  
V(x)-\e_l+a\}\,,            \nonumber\\
\overline a_0(x)&=&\frac{h^2(x)}{a}\{-h^2(x)-h^\prime(x)+ 
V(x)-\e_l+a\}                \nonumber\\
        & &+h^2(x)+h^\prime(x)-2\,h^\prime(x)-a\,.
              \nonumber
\ea
Therefore, if the function $h(x)$ satisfies the Riccati equation
$w^2+w^\prime=V(x)-\e_k$, with $\e_k=\e_l-a$,
and we rename it as $h(x)=w_k(x)$,
the new coefficients reduce to 
$\overline a_2(x)=-1$, $\overline a_1(x)=0$ and 
$\overline a_0(x)=V(x)-2\,w_k^\prime(x)-\e_l$.
\end{proof}

Let us note that in \cite{FerHusMiel98} the proof of
the Theorem~\ref{FerHusMiel_theorem} was just sketched. 
In addition, there exists an alternative proof; see, e.g.,
Mielnik, Nieto and Rosas-Ortiz \cite{mnro00}.

\section{Group elements preserving the subset of Riccati equations
$\lowercase{w^\prime+w^2}=V\lowercase{(x)-\e}$}

We have just seen how the transformation group defined on the set of
Riccati equations provides a direct proof of Theorem~\ref{FerHusMiel_theorem}. 
It relates one solution $w_l(x)$ of the \emph{initial} Riccati equation  
$w^\prime+w^2=V(x)-\e_l$ with one solution $w_{kl}(x)$ of the \emph{final} Riccati
equation $w^\prime+w^2=V(x)-2\,w_k^\prime(x)-\e_l$ by using a
solution $w_k(x)$ of the \emph{intermediate} Riccati equation
$w^\prime+w^2=V(x)-\e_k$. 
These three Riccati equations can be obtained from another three
Schr\"odinger-like equations by means of one of the reduction possibilities
explained in Section~3. Moreover, those associated with the initial and 
intermediate Riccati equations, namely $-\psi^{\prime\prime}+(V(x)-\e_l)\psi=0$
and $-\psi^{\prime\prime}+(V(x)-\e_k)\psi=0$, can be seen as the arising eigenvalue
equations for the two energies $\e_l$, $\e_k$ of the same potential $V(x)$, 
meanwhile the final Riccati equation can be associated to the eigenvalue
equation for the potential $V(x)-2\,w_k^\prime(x)$ with eigenvalue $\e_l$.

Then, we are naturally led to the question of which are 
the most general elements of $\GR$ preserving the subset of Riccati
equations characterized by having the coefficients 
$a_2(x)=-1$, $a_1(x)=0$ and $a_0(x)$ equal to some function, which 
we will write as an expression of the form  $V(x)-\e$. 

The Riccati equation we will start from is
\begin{equation}
w^\prime=-w^2+V(x)-\e\,,
\label{eq_u_gener}
\end{equation}
which according to (\ref{Riceq}) has the coefficients $a_2(x)=-1$,
$a_1(x)=0$ and $a_0(x)=V(x)-\e$. The condition for obtaining a final Riccati
equation in the mentioned subset is
\bea
\matriz{c}{-1 \\ 0 \\ \overline V(x)-\overline {\e}} 
&=&\matriz{ccc}
{{\d}^2&{-\d\g}&{{\g}^2} \\ {-2\,
\b\d}&{\a\d+\b\g}&{-2\,\a\g}\\{{\b}^2}&{-\a\b}&{{\a}^2}}
\matriz{c}{-1 \\ 0 \\ V(x)-\e} 
\nonumber\\     
& &+\matriz{c}{\g {{\d}^\prime}-\d {\g}^\prime\\
\d {\a}^\prime-\a {\d}^\prime+\b {\g}^\prime-\g {\b}^\prime\\
 \a {\b}^\prime-\b {\a}^\prime}\ ,
\label{transf_pots_matr}
\eea
for an $A(x)\in\GR$ given by (\ref{Agauge}) to be determined, and
where $\overline V(x)-\overline \e$ will be in general different to
$V(x)-\e$. Therefore, the elements of the subset of $\GR$ we are trying to
characterize will \emph{not} necessarily form a subgroup.
The matrix equation (\ref{transf_pots_matr}) is equivalent to
three scalar equations
\bea
-1&=&-{\d}^2+{\g}^2\,(V(x)-\e)+\g {{\d}^\prime}-\d {\g}^\prime\,,                
\label{eqesc1}\\
0&=&2\,\b\d-2\,\a\g\,(V(x)-\e)
+\d {\a}^\prime-\a {\d}^\prime+\b {\g}^\prime-\g {\b}^\prime\,,         
\label{eqesc2}\\
\overline V(x)-\overline \e
&=&-{\b}^2+{\a}^2\,(V(x)-\e)+\a{\b}^\prime-\b{\a}^\prime \,. 
\label{eqesc3}
\eea
Differentiating $\det A(x)=\a(x)\d(x)-\b(x)\g(x)=1$ we have as well
\begin{equation}
\a^\prime\d+\d^\prime\a-\g^\prime\b-\b^\prime\g=0\,.
\label{eqesc4}
\end{equation}
Out of these four equations, (\ref{eqesc1}), (\ref{eqesc2}) and
(\ref{eqesc4})  will give conditions on the matrix elements $\a$, $\b$,
$\g$, $\d$ and their derivatives such that the preserving condition be
satisfied.  The remaining (\ref{eqesc3}) will define 
$\overline V(x)-\overline \e$ in terms of all the other 
functions, including $V(x)-\e$. 

After taking the sum and the difference of (\ref{eqesc2}) and
(\ref{eqesc4}) it follows
\ba
(V(x)-\e)\a^2&=&\frac{\a\b\d}{\g}+\frac{\d\a\a^\prime}{\g}-\a\b^\prime\,,        
\label{Vc_grieg1}\\
(V(x)-\e)\g^2&=&\frac{\g\b\d}{\a}+\frac{\b\g\g^\prime}{\a}-\g\d^\prime\,.        
\label{Vc_grieg2}
\ea
Substituting them into (\ref{eqesc1}) and (\ref{eqesc3}) gives
\bea
-1&=&-{\d}^2+\frac{\g\b\d}{\a}+\frac{\b\g\g^\prime}{\a}-\d\g^\prime\,,  
\nonumber\\
\overline  V(x)-\overline \e&=&-{\b}^2+\frac{\a\b\d}{\g}
+\frac{\d\a\a^\prime}{\g}-\b\a^\prime\,.    \nonumber
\eea
Multiplying the first of these equations by $\a$ and the second by $\g$,
and using the fact that $\a\d-\b\g=1$, we arrive to
\ba
\a&=&\d+\g^\prime\,,                    \label{alfa_con_dgprime}        \\
(\overline  V(x)-\overline \e)\g&=&\b+\a^\prime\,.  
\label{Vtilctil_con_grieg}
\ea
Substituting (\ref{alfa_con_dgprime}) into (\ref{Vc_grieg2}) yields
\begin{equation}
(V(x)-\e)\g=\b-\d^\prime\,.      \label{Vc_con_grieg}
\end{equation}
We have two relations amongst the functions $\a$, $\b$, $\g$ and $\d$,
namely (\ref{alfa_con_dgprime}) and the determinant condition, so we can
express these matrix elements in terms of only two of them and their
derivatives. Then we have $\a=\d+\g^\prime$ and
$\b={(\d(\d+\g^\prime)-1)}/{\g}$.  Using moreover the fact that
\begin{equation}
\frac{\d^\prime}{\g}=\bigg(\frac{\d}{\g}\bigg)^\prime +
\frac{\d\g^\prime}{\g^2}
\nonumber
\end{equation}
the equation (\ref{Vc_con_grieg}) becomes
\begin{equation}
\bigg(-\frac{\d}{\g}\bigg)^\prime+\bigg(-\frac{\d}{\g}\bigg)^2
=V(x) +\frac 1 {\g^2}- \e \nonumber
\end{equation}
so the new function $v$ defined as $v=-{\d}/{\g}$ must satisfy the 
Riccati equation   
\begin{equation}
v^\prime+v^2=V(x) + \frac 1 {\g^2}-\e \,.\label{eq_Ric_v}
\end{equation}
Now, substituting in (\ref{Vtilctil_con_grieg}) the expressions of $\b$
and $\a^\prime$ in terms of $\d$, $\g$ and their derivatives, using the
definition of $v$ and the equation (\ref{eq_Ric_v}) gives
\begin{equation}
\overline V(x)-\overline \e
=V(x) - 2\left(\frac{\g^\prime}{\g}\,v+ v^\prime\right)
+\frac{\g^{\prime\prime}}{\g}-\e \,.      \nonumber
\end{equation}
It only remains to find the expression of the function solution of the
final Riccati equation, in terms of $w$ and $v$. The $SL(2,\R)$--valued
curve used for the transformation can be written as
\begin{equation}
C_0(x)=\g\,\matriz{cc}{{-v+\frac{\g^\prime}\g}
&{v^2-v\frac{\g^\prime}{\g}-\frac 1{\g^2}}\\{1}&{-v}}\,,
\label{matriz_transf}
\end{equation}
so the desired function is 
\ba
\overline w(x)&=&\Theta(C_0(x),w(x))=\frac{- v w + w
\g^\prime/\g-1/\g^2+v^2-v  
\g^\prime/\g}{w-v}   \nonumber\\
&=&-v-\frac{1/\g^2}{w-v}+\frac{\g^\prime}{\g}\,.
\ea
In short, we have just proved the following Theorem:

\begin{theorem} Let $w(x)$ be a solution of the Riccati equation 
\begin{equation} 
w^\prime+w^2= V(x) - \e 
\label{equtheorem}
\end{equation}
for some function $V(x)$ and some constant $\e$, and $\g(x)$ a never
vanishing differentiable function defined on the domain of\ $V(x)$.  
If $v(x)$ is a solution of the Riccati equation
\begin{equation}
v^\prime+v^2=V(x) + \frac{1}{\g^2(x)}-\e \,, 
\label{eqvtheorem}
\end{equation}
such that is defined in the same domain as $w(x)$ and $w(x)-v(x)$ does not vanish,
then the function $\overline w(x)$ defined
by
\begin{equation}
\overline w
(x)=-v(x)-\frac{1/\g^2(x)}{w(x)-v(x)}+\frac{\g^\prime(x)}{\g(x)}
\label{defztheor}
\end{equation}
is a solution of the Riccati equation 
\begin{equation}
\overline w^\prime+\overline w^2
=V(x) -
2\left(\frac{\g^\prime}{\g}\,v+v^\prime
\right)+\frac{\g^{\prime\prime}}{\g}- \e \,. 
\label{eqztheorem_1}
\end{equation}
\label{my_theor1}
\end{theorem}
Needless to say, the coefficients of the final equation can be calculated
directly by using (\ref{ta2}), (\ref{ta1}), (\ref{ta0}) and taking into
account (\ref{matriz_transf}), (\ref{equtheorem}) and (\ref{eqvtheorem}). 

\begin{corollary}
The Theorem~\ref{FerHusMiel_theorem} is a particular 
case of Theorem~\ref{my_theor1}.
\end{corollary}

\begin{proof}
It is sufficient to choose in Theorem~\ref{my_theor1} 
$w(x)=w_l(x)$, $v(x)=w_k(x)$, $\e=\e_l$ and $\g=1/\sqrt{\e_l-\e_k}$, with
$\e_k<\e_l$.
\end{proof}

Theorem~\ref{my_theor1} has a counterpart for linear second-order
differential equations of Schr\"odinger type, which will be in turn of
direct interest in physical applications. The key is to use in a inverse
way the reduction procedure outlined in Section~3. 

Consider the solution $w$ of the Riccati equation (\ref{equtheorem}).  We
can define (locally and up to a non-vanishing multiplicative constant)  
the new variable $\p_w$ as
\begin{equation}
\p_w(x)=\exp\bigg(\int^x w(\xi)\,d\xi\bigg)\,,
\label{def_p_u}
\end{equation}
which will satisfy 
\begin{equation}
-\p_w^{\prime\prime}+(V(x)-\e)\p_w=0\,,
\nonumber
\end{equation}
for the \emph{specific} constant $\e$.  Analogously, by considering a
solution $v$ of the Riccati equation (\ref{eqvtheorem}) we can define
(locally etc.) $\p_v$ as
\begin{equation}
\p_v(x)=\exp\bigg(\int^x v(\xi)\,d\xi\bigg)\,,
\label{def_p_v}
\end{equation}
which will satisfy 
\begin{equation}
-\p_v^{\prime\prime}+\bigg(V(x)+\frac 1 {\g^2(x)}-\e\bigg)\p_v=0\,,
\nonumber
\end{equation}
for the same specific constant $\e$. Then the function $\overline w$
defined by (\ref{defztheor}) will satisfy the Riccati equation
(\ref{eqztheorem_1}).  We could define as well (locally etc.) the new
function $\p_{\overline w}$ as
\begin{equation}
\p_{\overline w}(x)=\exp\bigg(\int^x {\overline w}(\xi)\,d\xi\bigg)\,,
\label{def_p_z}
\end{equation}
which in turn will satisfy 
\begin{equation}
-\p_{\overline w}^{\prime\prime}
+\bigg\{V(x) - 2\bigg(\frac{\g^\prime}{\g}\,v+v^\prime\bigg)
+\frac{\g^{\prime\prime}}{\g}- \e \bigg\}\p_{\overline w}=0\,.
\nonumber
\end{equation}
What has to be done now is to relate the function $\p_{\overline w}$ with
$\p_w$ and $\p_v$, taking into account the relation amongst ${\overline
w}$, $w$ and $v$. 

\begin{proposition} Let $w$, $v$, $\overline w$ be the functions for
which the Theorem~\ref{my_theor1} holds, and $\p_w$, $\p_v$,
$\p_{\overline w}$ the ones defined by (\ref{def_p_u}), (\ref{def_p_v}) 
and (\ref{def_p_z}), respectively. Then we have
\begin{equation}
\frac{\p_w^\prime}{\p_w}=w\,,
\quad\frac{\p_v^\prime}{\p_v}=v\,,
\quad\frac{\p_{\overline w}^\prime}{\p_{\overline w}}={\overline w}\,,
\end{equation}
and it holds 
\begin{equation}
\p_{\overline w}=\g\bigg(-\frac d
{dx}+\frac{\p_v^\prime}{\p_v}\bigg)\p_w\,,
\label{Adagen}
\end{equation}
up to a non-vanishing multiplicative constant.
\end{proposition} 

\begin{proof}
The first assertion is immediate. As a consequence, we have
$\g(-\frac d {dx}+\frac{\p_v^\prime}{\p_v})\p_w=\g(v-w)\p_w$.
Taking the logarithmic derivative
\ba
\frac{(\g(v-w)\p_w)^\prime}{\g(v-w)\p_w}
&=&\frac{\g^\prime}{\g}
+\frac{v^\prime-w^\prime}{v-w}+\frac{\p_w^\prime}{\p_w} \nonumber \\
&=&\frac{\g^\prime}{\g}+\frac{w^2-v^2}{v-w}+\frac{1/\g^2}{v-w}+w
=\frac{\g^\prime}{\g}-w-v+\frac{1/\g^2}{v-w}+w   \nonumber\\
&=&\frac{\g^\prime}{\g}-v+\frac{1/\g^2}{v-w}=\overline 
w=\frac{\p_{\overline w}^\prime}{\p_{\overline w}}\,,  
\nonumber
\ea
where it has been used the equations (\ref{equtheorem}),
(\ref{eqvtheorem}) and (\ref{defztheor}). 
\end{proof}

With the previous results we have the following:

\begin{theorem} 
Let $\p_w(x)$ be a solution of the ho\-mo\-ge\-neous li\-near second order
differential equation
\begin{equation}
-\p_w^{\prime\prime}+(V(x)-\e)\p_w=0\,,
\label{Scho_u_theor}
\end{equation}
for some \emph{specific} function $V(x)$ and constant $\e$, and $\g(x)$ 
a never vanishing differentiable function defined on the domain of\
$V(x)$. If the function $\p_v(x)\neq\p_w(x)$ is a solution of the
equation
\begin{equation}
-\p_v^{\prime\prime}+\bigg(V(x)+\frac 1 {\g^2(x)}-\e\bigg)\p_v=0\,,
\label{Scho_v_theor}
\end{equation}
defined in the same domain as $\p_w(x)$, then the
function $\p_{\overline w}(x)$ defined (up to a non-vanishing
multiplicative constant) by
\ba
\p_{\overline w}=\g\left(-\frac d
{dx}+\frac{\p_v^\prime}{\p_v}\right)\p_w\,,
\label{Adagen_theor}
\ea
satisfies the new equation
\ba
-\p_{\overline w}^{\prime\prime}
+\left\{V(x) 
- 2\bigg(\frac{\g^\prime}{\g}\,v+v^\prime\bigg)
+\frac{\g^{\prime\prime}}{\g} - \e \right\}\p_{\overline w}=0\,,
\label{Scho_z1_theor}
\ea
where the function $v(x)$ is defined (locally) as $\p_v^\prime/\p_v=v$.
\label{my_theor2}
\end{theorem}

Note that Theorems~\ref{my_theor1} and 
\ref{my_theor2} are invariant under the change of sign of $\g$.

\section{Finite difference algorithm and intertwined Hamiltonians from a
group theoretical viewpoint}

We have already said that the finite difference algorithm, based on 
the Theorem~\ref{FerHusMiel_theorem}, appeared in \cite{FerHusMiel98}
when the authors wanted to \emph{iterate} the standard first order
intertwining technique. This idea has been kept also in subsequent
works \cite{FerHus99,Ros}, and in all of these articles the algorithm 
has been shown to be of use for obtaining new exactly solvable Hamiltonians.
Moreover, the proof of Theorem~\ref{FerHusMiel_theorem} 
given recently by Mielnik, Nieto and Rosas-Ortiz, alternative 
to that which has been given here, still relies on the idea of iteration
of the intertwining technique, see \cite[Sec. 2]{mnro00} for details. 

On the other hand, we have given a direct proof of Theorem~\ref{FerHusMiel_theorem}
by making use of the action of $\GR$ on the set of Riccati equations, and 
we wonder whether it is possible to establish a further relation between 
this transformation group and the (maybe iterated) intertwining technique. 

The important result, which we show next, is the following.
By using properly 
the finite difference algorithm just \emph{once}, 
jointly with the reduction procedure described in Section~3, 
it is possible to explain from a group theoretical viewpoint 
the usual problem of $A$--related or intertwined Hamiltonians. 

With this aim, let us consider two Hamiltonians
\begin{equation}
H_0=-\frac{d^2}{dx^2}+V_0(x)\,,\quad\quad H_1=-\frac{d^2}{dx^2}+V_1(x)\,,
\label{HamiltoniansH0H1}
\end{equation}
which by hypothesis are $A_1$--related, i.e. $A_1 H_1=H_0 A_1$ and $H_1
A_1^\dagger=A_1^\dagger H_0$, where
\begin{equation}
A_1=\frac d {dx}+w_1\,,\quad\mbox{}\quad A_1^\dagger=-\frac d
{dx}+w_1\,,
\end{equation}
and $w_1$ is a function to be determined. 

Assume that $H_0$ is an exactly solvable Hamiltonian for which we know a
complete set of square-integrable eigenfunctions $\ps^{(0)}_n$ with
respective energies $E_n$, $n=0,\,1,\,2,\,\dots$. We have seen in
Section~3 that, in particular,
\ba
V_0(x)-E_0&=&w_1^2(x,E_0)+w_1^\prime(x,E_0)\,,   \label{P1_2}\\
V_1(x)-E_0&=&w_1^2(x,E_0)-w_1^\prime(x,E_0)\,,   \label{P1_1}
\ea
or equivalently 
\ba
V_0(x)-E_0&=&-(V_1(x)-E_0)+2\,w_1^2(x,E_0)\,,   \label{P2_1}\\
V_1(x)&=&V_0(x)-2\,w_1^\prime(x,E_0)\,. \label{P2_2}
\ea
where we have chosen $w_1(x,E_0)$ as
\begin{equation}
w_1(x,E_0)={\,\ps_0^{(0)\prime}}/{\ps_0^{(0)}}\,. \label{defa_1E0}
\end{equation}
Up to a non-vanishing multiplicative constant, we define the function
$\ps_0^{(1)}$ as $\ps_0^{(1)}=1/{\ps_0^{(0)}}$. We have as well
\begin{equation}
w_1(x,E_0)=-{\,\ps_0^{(1)\prime}}/{\ps_0^{(1)}}\,. \label{defa_1E0b}
\end{equation}
Then, both Hamiltonians factorize as
\begin{equation}
H_0=A_1(E_0)A_1^\dagger(E_0)+E_0\,,\quad\quad\quad 
H_1=A_1^\dagger(E_0)A_1(E_0)+E_0\,.
\label{factH0H1E0}
\end{equation}

We have made explicit $E_0$ in the function $w_1$ and, as a consequence, in
the operators $A_1$ and $A_1^\dagger$. However, 
it should be considered as a label reminding the factorization we are working with
rather than as a functional
dependence.  {}From (\ref{defa_1E0}) and
(\ref{defa_1E0b}) we have $A_1^\dagger(E_0)\ps_0^{(0)}=0$ and
$A_1(E_0)\ps_0^{(1)}=0$; as a result $H_1\ps_0^{(1)}=E_0\ps_0^{(1)}$ and
$H_0 \ps^{(0)}_0=E_0 \ps^{(0)}_0$.  As $\ps_0^{(0)}$ has no zeros in the
domain of $V_0(x)$, all the functions defined in this Section will be
globally defined provided that such a domain is connected. 

The equation (\ref{P2_2}) relates the new potential $V_1(x)$ and the old
one $V_0(x)$. As it is well known, due to the $A_1(E_0)$--relationship of
the Hamiltonians $H_0$ and $H_1$, the normalized eigenfunctions of $H_1$
can be obtained transforming those of $H_0$ by means of the operator
$A_1^\dagger(E_0)$ except $\ps^{(0)}_0$, since
$A_1^\dagger(E_0)\ps^{(0)}_0=0$. In fact, a simple calculation shows that
the functions
\begin{equation}
\ps_n^{(1)}=\frac{A_1^\dagger(E_0)\ps_n^{(0)}}{\sqrt{E_n-E_0}}\,,
\label{nweifn_FM}
\end{equation}
satisfy
\begin{equation}
H_1 \ps_n^{(1)}=E_n\ps_n^{(1)}\quad\quad\mbox{and}
\quad\quad(\ps_n^{(1)},\ps_m^{(1)})=\d_{nm}\,,
\end{equation}
for all $n,\,m=1,\,2,\,3,\,\dots$, provided that the functions $\ps_n^{(0)}$ are
normalized. 

Although the function $\ps_0^{(1)}$ satisfies 
$H_1\ps_0^{(1)}=E_0\ps_0^{(1)}$, it does not correspond to a physical
state of $H_1$ since it is not normalizable, which means that $E_0$ will
not belong to the spectrum of $H_1$. For this reason the Hamiltonians
$H_1$ and $H_0$ are said to be \emph{quasi-isospectral\/}.
 
Let us formulate now these results in terms of the transformation group on
the set of Riccati equations introduced in Section~4. By hypothesis we
have
\begin{equation}
H_0 \ps_n^{(0)}=E_n \ps_n^{(0)}\,,\quad\quad\quad n=0,\,1,\,2,\,\dots\,.
\label{H0solvable}
\end{equation}
As $H_0$ is given by (\ref{HamiltoniansH0H1}), the spectral equation
(\ref{H0solvable}) can be written as the following \emph{set} of equations
\begin{equation}
-\ps_n^{(0)\prime\prime}+(V_0(x)-E_n)\ps_n^{(0)}=0\,,
\quad\quad\quad n=0,\,1,\,2,\,\dots
\label{H0solvable_coord}
\end{equation}
We introduce the new functions
\begin{equation}
w_1(E_n)=\frac{\ps_n^{(0)\prime}}{\ps_n^{(0)}}\,,
\quad\quad\quad n=0,\,1,\,2,\,\dots\,,
\label{ch_Ek}
\end{equation}
where the dependence on $x$ has been omitted for brevity.  As we know from
Section~3, these transformations will be defined \emph{locally}, {\it
i.e.}, for each $n$ the domain of $w_1(E_n)$ will be the union of the open
intervals contained between two consecutive zeros of $\ps_n^{(0)}$ or
maybe a zero and one boundary of the domain of $V_0(x)$.  In particular,
$w_1(E_0)$ is defined globally in the entire domain of $V_0(x)$ as
$\ps_0^{(0)}$ have no zeros there. Therefore, the set of equations
(\ref{H0solvable_coord}) reads in the new variables as the set
\begin{equation}
w_1^\prime(E_n)+w_1^2(E_n)=V_0(x)-E_n\,,\quad\quad\quad
n=0,\,1,\,2,\,\dots\,,
\nonumber
\end{equation}
that is, the functions $w_1(E_n)$ are respective solutions of the
Riccati equations 
\begin{equation}
w^\prime+w^2=V_0(x)-E_n\,,\quad\quad\quad
n=0,\,1,\,2,\,\dots\,.
\label{H0solvable_coord_Ricc}
\end{equation}

Let us apply the Theorem~\ref{FerHusMiel_theorem} to this situation. 
We \emph{act} on the set of all equations (\ref{H0solvable_coord_Ricc}) 
but one by means of suitable group elements
of $\GR$. These $SL(2,\R)$--valued curves are constructed by means of 
the solution of the equation of the set (\ref{H0solvable_coord_Ricc}) which 
is to be set aside. In order to avoid singularities, this solution should 
be the one with $n=0$.

The mentioned elements of $\GR$ are analogous to the one used in the
proof of Theorem~\ref{FerHusMiel_theorem}. They turn out to be 
\begin{equation}
B_n=\frac{1}{\sqrt{E_n-E_0}}
\matriz{cc}{{w_1(E_0)}&{-w_1^2(E_0)+E_n-E_0}\\{-1}&{w_1(E_0)}}\,,
\quad n=1,\,2,\,\dots\,.
\label{B0inGR}
\end{equation}
We define the new functions $\overline w_1(E_n)$ by 
\ba
\overline w_1(E_n)&=&\Theta(B_n,w_1(E_n))
=\frac{w_1(E_0)w_1(E_n)-w_1^2(E_0)+E_n-E_0}{w_1(E_0)-w_1(E_n)}
\nonumber\\
&=&-w_1(E_0)-\frac{E_0-E_n}{w_1(E_0)-w_1(E_n)}\,,
\quad\quad n=1,\,2,\,\dots\,.           \label{definition_bEn}
\ea
By Theorem~\ref{FerHusMiel_theorem} these functions satisfy, respectively, 
the new Riccati equations
\begin{equation}
\overline w^\prime+\overline w^2
=V_1(x)-E_n\,,\quad\quad\quad n=1,\,2,\,\dots\,,
\label{Ricc_betas_transform}
\end{equation}
where $V_1(x)=V_0(x)-2\,w_1^\prime(E_0)$.  We can define (locally etc.)
the new set of functions $\p_n^{(1)}$, for $n=1,\,2,\,\dots$, by
\begin{equation}
\p_n^{(1)}(x)=\exp\bigg(\int^x \overline w_1(\xi,E_n)\,d\xi\bigg)\,,
\quad\quad\quad n=1,\,2,\,\dots\,,
\label{def_phis_betas}
\end{equation}
which therefore satisfy, respectively, a  
linear second-order differential equation of the set
\begin{equation}
-\p^{\prime\prime}+(V_1(x)-E_n)\p=0\,,
\quad\quad\quad n=1,\,2,\,\dots\,.
\label{phis_eigenstates}
\end{equation}
Then, $\p_n^{(1)}$ are eigenfunctions of the Hamiltonian $H_1=-\frac{d^2}
{dx^2}+V_1(x)$ with associated eigenvalues $E_n$, for $n=1,\,2,\,\dots$.
As a consequence, they can be written as the linear combinations
$\p_n^{(1)}=\ps_n^{(1)}
+\lambda_n\,\ps_n^{(1)}(x)\int^x \frac{d\xi}{\ps_n^{(1)\,2}(\xi)}$, for all
$n=1,\,2,\,\dots\,$, up to non-vanishing constant factors, where $\lambda_n$
are still unknown constants and it has been used the well-known
Liouville formula for finding the second linearly independent solution of each
equation of (\ref{phis_eigenstates}) starting from $\ps_n^{(1)}$.

Now, the important point is that each of the functions $\p_n^{(1)}$ turns out to
be the same as $\ps_n^{(1)}$, up to a non-vanishing constant factor, i.e. the
previous constants $\lambda_n$ are all zero.
In fact, the logarithmic derivative of $\ps_n^{(1)}$ is
\ba
\frac{\,\ps_n^{(1)\prime}}{\ps_n^{(1)}}
&=&\frac{\frac{d}{dx}(A_1^\dagger(E_0)\ps_{n}^{(0)})}
{A_1^\dagger(E_0)\ps_{n}^{(0)}}                         
=\frac{\frac{d}{dx}(-\ps_n^{(0)\prime}
+\frac{\ps_0^{(0)\prime}}{\ps_0^{(0)}}\,\ps_n^{(0)})}
{-\ps_n^{(0)\prime}+\frac{\ps_0^{(0)\prime}}{\ps_0^{(0)}}\,\ps_n^{(0)}} 
\nonumber\\
&=&\frac{
-\ps_{n}^{(0)\prime\prime}
+\frac{\ps_{0}^{(0)\prime\prime}}{\ps_{0}^{(0)}}\,\ps_{n}^{(0)}
-\bigg(\frac{\ps_{0}^{(0)\prime}}{\ps_{0}^{(0)}}\bigg)^2\,\ps_{n}^{(0)}
+\frac{\ps_{0}^{(0)\prime}}{\ps_{0}^{(0)}}\,\ps_{n}^{(0)\prime}}
{-\ps_{n}^{(0)\prime}+\frac{\ps_{0}^{(0)\prime}}{\ps_{0}^{(0)}}\, 
\ps_{n}^{(0)}}\,. 
                                                        \nonumber
\ea
Taking common factor $\ps_{n}^{(0)}$ in both numerator and denominator, using 
the relations
\begin{equation}
\frac{\ps_{n}^{(0)\prime\prime}}{\ps_{n}^{(0)}}=V_0(x)-E_n\,,
\quad\quad n=0,\,1,\,2,\,\dots\,,       \nonumber
\end{equation}
and the definitions (\ref{ch_Ek}), we arrive to
\begin{equation}
\frac{\,\ps_n^{(1)\prime}}{\ps_n^{(1)}}
=-w_1(E_0)-\frac{E_0-E_n}{w_1(E_0)-w_1(E_n)}
=\overline w_1(E_n)=\frac{\p_n^{(1)\prime}}{\p_n^{(1)}}\,,
\end{equation}
for $n=1,\,2,\,\dots\,$, therefore $\ps_n^{(1)}$ and $\p_n^{(1)}$ must
be proportional. It is also clear that these equations hold interval-wise. 

Now, as far as $\ps_0^{(0)}$ is concerned, it is clear that
Theorem~\ref{FerHusMiel_theorem} would make no sense for $w_k(x)=w_l(x)$
and $\e_k=\e_l$. In a similar way, we cannot put $E_n=E_0$ in
(\ref{B0inGR}):  the normalizing factors ${1}/{\sqrt{E_n-E_0}}$, which
were introduced in order to get $SL(2,\R)$--valued curves, would no longer make
sense because these matrices, after dropping out such factors would have
zero determinant. This means that one \emph{cannot} use a transformation
of type (\ref{definition_bEn}) with $E_n=E_0$ for the function $w_1(E_0)$
itself. However, the associated function to $\ps_0^{(1)}$ at the Riccati level
is just given by (\ref{defa_1E0b}), i.e. the new function 
$\overline w_1(E_0)=-w_1(E_0)$ satisfies an equation of type
(\ref{Ricc_betas_transform}) \emph{for $n=0$}, which is precisely
(\ref{P1_1}). In short, the equation $A_1^\dagger(E_0)\ps^{(0)}_0=0$ is
translated at the Riccati level into the fact that $w_1(E_0)$ cannot be
transformed in the sense mentioned above. Conversely, it is not
possible to write $\overline w_1(E_0)=\Theta(B_0,w_1(E_0))$ for
$B_0\in\GR$.
 
We have just explained the problem of two $A_1(E_0)$--related Hamiltonians
by means of Theorem~\ref{FerHusMiel_theorem}, which in turn is a
particular case of Theorem~\ref{my_theor1}. For the sake of completeness,
let us show briefly how Theorem~\ref{my_theor2} applies to the same
problem. Consider the set of equations (\ref{H0solvable_coord}), where we
retain again the one with $n=0$, which will play the r\^ole of equation
(\ref{Scho_v_theor}). All the others will play the r\^ole of equation
(\ref{Scho_u_theor}).  For each $n=1,\,2,\,\dots$, the function $\g$ must
be defined by
$$
-E_0=-E_n+\frac 1 {\g^2}\,.
$$
{}Thus, we can choose $\g=1/\sqrt{E_n-E_0}$. According to
(\ref{Adagen_theor}) and (\ref{Scho_z1_theor}) the functions
\begin{equation}
\vp_n^{(1)}=\frac 1{\sqrt{E_n-E_0}}
\left(-\frac d {dx}+\frac{\ps_0^{(0)\prime}}{\ps_0^{(0)}}\right)\ps_n^{(0)}\,,
\quad\quad\quad n=1,\,2,\,\dots\,
\label{trans_map_variable}
\end{equation}
satisfy, respectively, 
$$
-\vp^{\prime\prime}+(V_0(x)-2\,w_1^\prime(E_0)-E_n)\vp=0\,,
\quad\quad\quad n=1,\,2,\,\dots\,,
$$
where it has been used $w_1(E_0)={\ps_0^{(0)\prime}}/{\ps_0^{(0)}}$.  
In this way we have recovered the normalized eigenfunctions (\ref{nweifn_FM})
associated to the new potential $V_1(x)=V_0(x)-2\,w_1^\prime(x,E_0)$.

\section{Illustrative examples}

In this Section we will apply Theorem~\ref{my_theor2} to some cases where
$\g(x)$ is not a constant, what provides a more general situation than
the usual intertwining technique. However, note that with this method
we will be able to find potentials for which \emph{one} 
eigenfunction and its corresponding eigenvalue
are exactly known. We will use a slight generalization of two
well-known types of potentials, namely the radial oscillator and Coulomb
potentials, which consists of taking the most general
intervals of the appearing parameters such that
it is possible to find square-integrable eigenfunctions. 

\subsection{Radial oscillator-like potentials\label{rad_osc_like}}

Let us consider the family of potentials 
\begin{equation}
V_{l,b}(x)=\frac{b^2 x^2}{4}+\frac{l(l+1)}{x^2}\,,
\label{pot_oscil1}
\end{equation}
where $x\in(0,\infty)$ and $l$, $b$ are real parameters.  Each member can
be regarded as being part of a pair of shape invariant partner potentials,
with associated transformation law $l\rightarrow l+1$ for the parameter
$l$ \cite{InfHul51,CarRam2}.  
This fact allows to find the eigenvalues and the corresponding
eigenfunctions, even normalized, in an algebraic way.  The key is to find
functions in the kernel of the first order differential operator
$d/dx-(l+1)/x+b x/2$ which be normalizable with respect to the norm
induced by the standard scalar product 
defined in $L^2(0,\infty)$. That
will provide the ground state eigenfunction. 
The eigenfunctions of the excited states are
obtained {}from the iterated action of $-d/dx-(l+1)/x+b x/2$, with
appropriate $l$ at each step, times some suitable factors, on the ground state
eigenfunction. However, the point is that with this procedure, one obtains
the boundary conditions of the eigenfunctions as a consequence rather than
being {\it a priori} requirements. The result for this family of
potentials are the normalized eigenfunctions (up to a modulus one factor)
\cite{CarRam4}
\begin{equation}
\z_k^{l,b}(x)=\sqrt{\frac{\Gamma(k+1)}{\Gamma(k+l+3/2)}}
\bigg(\frac{b^{2 l+3}}{2^{2l+1}}\bigg)^{1/4} 
x^{l+1} e^{-b x^2/4} L_k^{l+1/2}\bigg(\frac {b x^2}2\bigg)\,,
\label{eigfunc_pot_oscil}
\end{equation}
where $k=0,\,1,\,2,\,\dots$, for $l>-3/2$ and $b>0$. The notation
$L_n^a(u)$ means the Laguerre polynomial of degree $n$ and parameter $a$
in the variable $u$. 

Note that in the interval $l\in(-3/2,-1)$ these eigenfunctions
go to infinity as $x$ tends to zero, contrary to the usual
requirement of going to zero, in spite of the fact of being 
square-integrable. 

The problem of the quantum-mechanical motion of a particle in a 
potential on the half line $(0,\infty)$
has been carefully studied in \cite{RedSim75}. 
It has been shown there that the operator 
$H=-\frac{d^2}{dx^2}+V(x)$, with domain
$C_0^\infty(0,\infty)$ of differentiable 
functions with compact support in  
$(0,\infty)$,
$V$ being a continuous real-valued function on $(0,\infty)$,  
is a symmetric operator and that 
it is essentially self-adjoint if and only if 
$V(x)$ is in the limit point case in both zero and infinity 
\cite[Theorem {\bf X.7}]{RedSim75}. 
In the case we have in hand, 
what happens is that the potentials
of the family (\ref{pot_oscil1}) lead to 
essentially self-adjoint 
Hamiltonians $-d^2/dx^2+V_{l,b}(x)$ for the range 
$b>0$ and $l>-3/2$, 
with different self-adjoint extensions in each of 
the intervals $l\in(-3/2,-1)$ and $l\in(-1,\infty)$, 
the first including
eigenfunctions which do not necessarily go to 
zero as $x\rightarrow 0$.
We will see that one eigenfunction arising when 
$l\in(-3/2,-1)$ provides an interesting application, 
for the family of potentials (\ref{pot_oscil1}), 
of our new method generalizing the first order intertwining technique.

In both cases, the corresponding eigenvalues to the eigenfunctions  
(\ref{eigfunc_pot_oscil}) for the potentials (\ref{pot_oscil1}) 
are \cite{CarRam4}:
\begin{equation}
E_k^{l,b}=b\bigg(2 k+l+\frac 3 2\bigg)\,, \qquad k=0,\,1,\,2,\,\dots\,.
\end{equation}
If $b=2$ these eigenvalues reduce to those of \cite{FerNegOlm}. Compare
also with the eigenfunctions and eigenvalues given in
\cite[pp. 391--2]{BagSam97}.

\begin{example}
\label{ex_oscil_inter}
Let us consider the following variant of the family 
of potentials (\ref{pot_oscil1})
\begin{equation}
V_{l,b}(x)=\frac{b^2 x^2}{4}+\frac{l(l+1)}{x^2}-b\bigg(l+\frac 3 2\bigg)\,,
\label{pot_oscil2}
\end{equation}
where $x\in(0,\infty)$, $l>-3/2$ and $b>0$. The normalized eigenfunctions
are given again by (\ref{eigfunc_pot_oscil}), with the same peculiarities,
but the corresponding eigenvalues are now: 
\begin{equation}
E_k^{l,b}=2 b k\,, \qquad k=0,\,1,\,2,\,\dots\,.
\end{equation}

We would like to apply Theorem~\ref{my_theor2}, by using
two potentials of the family (\ref{pot_oscil2}) for different
\emph{specific} values of $l$. The difference between them should be a
positive function in $(0,\infty)$ in order to define
appropriately $\g(x)$ as required by the Theorem. We have
$$
V_{l,b}(x)-V_{l+r,b}(x)=r\bigg(b-\frac{2l+1+r}{x^2}\bigg)\,,
$$
where $r$ is a positive integer. Since $b>0$, the condition for the right
hand side to be always positive is that $2l+1+r<0$. We can find a solution
if $r=1$, since then it should happen $2l+2<0$ or equivalently $l<-1$. For
$r=2,\,3,\,\dots$, we would find $l<-3/2$, which is incompatible with the
range of $l$. Then, we have to choose $r=1$, $-3/2<l<-1$ and 
therefore an appropriate function $\g(x)$ is
$$
\g_{l,b}(x)=\bigg(b-\frac{2l+2}{x^2}\bigg)^{-1/2}\,.
$$
We will transform an eigenstate of $V_{l+1,b}(x)$ by making use
of the eigenstate of $V_{l,b}(x)$ with the same energy, 
i.e. with the same $k$. Consider the functions 
$$
v_k^{l,b}(x)=\frac{1}{\z_k^{l,b}(x)}\frac{d \z_k^{l,b}(x)}{d x}\,,
$$
one of which will be used to find the final potential according 
to (\ref{Scho_z1_theor}). Due to the presence of the 
Laguerre polynomials in (\ref{eigfunc_pot_oscil}), 
$\z_k^{l,b}(x)$ has $k$ zeros in $(0,\infty)$ and therefore
$v_k^{l,b}(x)$, as well as the final potential, have $k$ 
singularities in the same interval. In order to avoid them,
we choose $k=0$.
Summarizing, we transform the
eigenfunction $\z_0^{l+1,b}(x)$ obeying
\begin{equation}
-\frac{d^2 \z_0^{l+1,b}(x)}{dx^2}+V_{l+1,b}(x)\z_0^{l+1,b}(x)=0\,,
\label{eq_ej_osc2}
\end{equation}
by means of the solution $\z_0^{l,b}(x)$ of a equation similar to
(\ref{eq_ej_osc2}) but with $l$ instead of $l+1$. 
Since $l\in(-3/2,-1)$, both of the original eigenfunction $\z_0^{l+1,b}(x)$ 
and the intermediate one $\z_0^{l,b}(x)$ are square-integrable, but
this last goes to infinity when $x\rightarrow 0$.

After some calculations, the image potential becomes
\ba
V^{im}_{l,b}(x)&=&
V_{l+1,b}(x)-2\bigg(\frac{\g_{l,b}^{\prime}}{\g_{l,b}}v_0^{l,b}
+\frac {d v_0^{l,b}}{dx}\bigg)+\frac{\g_{l,b}^{\prime\prime}}{\g_{l,b}} 
\nonumber\\
&=&\frac{b^2 x^2}{4}+\frac{(l+1)(l+2)}{x^2}-b\bigg(l+\frac 3 2\bigg)
+\frac{6 b(l+1)}{(b x^2-2(l+1))^2}\,,
\nonumber
\ea
for which we obtain the eigenstate with zero energy
\ba
\eta_0^{l,b}(x)&=&
\g_{l,b}(x)\bigg(-\frac{d\z_0^{l+1,b}(x)}{dx}+v_0^{l,b}\z_0^{l+1,b}(x)\bigg)
\nonumber\\
&=&\sqrt{\frac{b^{l+5/2}}{2^{l+3/2} \Gamma(l+5/2)}}
\frac{x^{l+2}e^{-b x^2/4}}{\sqrt{b x^2-2(l+1)}}\,,
\nonumber
\ea
as can be checked by direct calculation. Notice that $b x^2-2(l+1)>0$
always since $l<-1$ and $b>0$, and therefore 
$V^{im}_{l,b}(x)$ and $\eta_0^{l,b}(x)$ are defined in the whole 
interval $(0,\infty)$. Moreover,
$\eta_0^{l,b}(x)$ has no zeros, and it tends to zero when $x$ goes to $0$
and to $\infty$ fast enough to give a square-integrable eigenfunction.  In
fact, it can be easily checked that
$$
(\eta_0^{l,b},\eta_0^{l,b})=\int_0^\infty |\eta_0^{l,b}(x)|^2\,dx
=\frac{e^{-l-1}}{2}(-l-1)^{l+3/2}\Gamma\bigg(-l-\frac 3 2,-l-1\bigg)\,,
$$  
where $\Gamma(\a,x)$ denotes the incomplete Gamma function defined by
$\Gamma(\a,x)=\int_x^\infty e^{-t} t^{\a-1}\,dt$. The previous formula can
be derived by means of the change of variable $bx^2=2t$ and using
\cite[Formula 8.353.3]{GraRyz}:
$$
\Gamma(\a,x)=\frac{e^{-x}x^\a}{\Gamma(1-\a)}
\int_0^\infty\frac{e^{-t}t^{-\a}}{t+x}\,dt\,,\quad\quad \mbox{Re}\,\a<1,\,x>0\,.
$$
As we can see, the norm of the final eigenfunction depends on $l$, not on $b$, 
and it takes real values only if $l<-1$, in agreement with the range of
application for $l$ previously derived.
\end{example}

\subsection{Radial Coulomb-like potentials}

Let us consider now the family of potentials 
\begin{equation}
V_{l,q}(x)=\frac{2 q}{x}+\frac{l(l+1)}{x^2}\,,
\label{pot_Coul_1}
\end{equation}
where $x\in(0,\infty)$ and  $q\neq 0$, $l$ are real parameters. 
This family shares several characteristics with that of 
(\ref{pot_oscil1}). For a start, each member can be regarded as 
being part of a pair of shape invariant partner potentials, respect to
the transformation law $l\rightarrow l+1$ \cite{InfHul51,CarRam2}. 
Similarly as before, one can
obtain the normalized eigenfunctions 
(up to a modulus one factor) \cite{CarRam4} 
\begin{equation}
\z_k^{l,q}(x)=\sqrt{\frac{\Gamma(k+1)}{\Gamma(2l+2+k)}}
\frac{2^{l+1}|q|^{l+3/2}}{(k+l+1)^{l+2}}\,
x^{l+1} e^{\frac{q x}{k+l+1}} L_k^{2 l+1}\bigg(\frac {-2 q x}{k+l+1}\bigg)\,.
\label{eigfunc_pot_Coul}
\end{equation} 
These eigenfunctions are square-integrable only in the following circumstances: 
for $l\in(-3/2,-1)$ and $q>0$, only the eigenfunction with $k=0$. For
$l\in(-1,\infty)$ and $q<0$, the functions (\ref{eigfunc_pot_Coul}) are
normalizable for all $k=0,\,1,\,2,\,\dots$. The normalizable solution in the
range $l\in(-3/2,-1)$, $q>0$, goes to infinity as $x$ tends to zero,
meanwhile all the others go to zero as $x\rightarrow 0$. Again, the reason
is the existence of different self-adjoint extensions on the different ranges,
being the Hamiltonians $H_{l,q}=-\frac{d^2}{dx^2}+V_{l,q}(x)$ essentially
self-adjoint if $l>-3/2$ and $q/(l+1)<0$.

The corresponding eigenvalues to the eigenfunctions 
(\ref{eigfunc_pot_Coul}) for the family (\ref{pot_Coul_1}) 
are \cite{CarRam4}:
\begin{equation}
E_k^{l,q}=-\frac{q^2}{(k+l+1)^2}\,, \qquad k=0,\,1,\,2,\,\dots
\end{equation}
If $q=-1$ and thus $l>-1$ we recover the spectrum given, for example, 
in \cite{Ros}. Compare also with \cite[p. 389]{BagSam97}.

\begin{example}
\label{coul_norm}
Let us use now Theorem~\ref{my_theor2} with two potentials of the
family (\ref{pot_Coul_1}) with different values of $l$.
We ask that
$$
V_{l,q}(x)-V_{l-r,q}(x)=\frac{(2l+1-r)r}{x^2}=\frac{1}{\g_{l,r}^2(x)}\,,
$$
where $r>0$ is to be determined below, 
so we can choose $\g_{l,r}(x)=x/\sqrt{r(2l+1-r)}$. 
We will transform one eigenfunction $\z_k^{l-r,q}(x)$ which satisfies
$$
-\frac{d^2 \z_k^{l-r,q}(x)}{d x^2}
+\bigg\{V_{l-r,q}(x)+\frac{q^2}{(k+l-r+1)^2}\bigg\}\z_k^{l-r,q}(x)=0\,,
$$
for some $k=0,\,1,\,2,\,\dots\,,$ by using some suitable solution
of the equation 
$$
-\frac{d^2 \p_v}{d x^2}
+\bigg\{V_{l,q}(x)+\frac{q^2}{(k+l-r+1)^2}\bigg\}\p_v=0\,.
$$
A natural idea is to choose $\p_v(x)$ as one of the 
eigenfunctions $\z_m^{l,q}(x)$ of $V_{l,q}(x)$ 
for certain integer $m$ defined by the condition
$$
E_m^{l,q}=-\frac{q^2}{(k+l-r+1)^2}\,,
$$
whose simplest solution is
$m=k-r$. Since $m$ and $k$ are non-negative integers, we have
$k\geq r$ and therefore $r$ must be a non-negative integer as well.  
As in Example~\ref{ex_oscil_inter}, in order to avoid
singularities in the final potential, we have to take $m=0$ and hence
$k=r$. Then, we transform the eigenfunction 
corresponding to the integer $k>0$ of the
potential $V_{l-k,q}(x)$, with eigenvalue $-q^2/(l+1)^2$, by using the ground
state of the potential $V_{l,q}(x)$, with the same energy eigenvalue. 
The original eigenfunction has to be normalizable, so it must be
$l-k>-1$ and hence $q<0$, because in the range $l\in(-3/2,-1)$ 
there are only one normalizable eigenfunction and $k>0$. 
Consequently, $l>k-1\geq 0$, and both of the initial and intermediate eigenfunctions
are square-integrable and go to zero as $x\rightarrow 0$.
If we denote $v_0^{l,q}(x)=(1/\z_0^{l,q}(x))d{\z_0^{l,q}}(x)/dx$, the image
potential reads
$$
V_{l-k,q}(x)-2\bigg(\frac{v_0^{l,q}}x+\frac {d v_0^{l,q}}{dx}\bigg)
=V_{l-k,q}(x)-\frac{2q}{(l+1)x}=V_{l-k,q\,l/(l+1)}(x)\,.
$$

Correspondingly we find, after some calculations, the final eigenfunction
$$ 
\eta_k^{l,q}(x)=
\g_{l,k}(x)\bigg(-\frac{d\z_k^{l-k,q}(x)}{dx}+v_0^{l,q}\z_k^{l-k,q}(x)\bigg) 
=\sqrt{\frac l{l+1}} \z_{k-1}^{l-k,q\,l/(l+1)}(x)\,.
$$ 
In this way we recover the original potential $V_{l-k,q}(x)$ but with the
coupling constant $q$ scaled by the factor $l/(l+1)>0$. This scaling is also
reflected in the final eigenfunction, which moreover has $k-1$ instead of $k$,
and norm $\sqrt{l/(l+1)}$. 
\end{example}

\begin{example}
\label{coul_variant_1}
We will consider now the following modified version of the potentials 
(\ref{pot_Coul_1}):
\begin{equation}
V_{l,q}(x)=\frac{2 q}{x}+\frac{l(l+1)}{x^2}+\frac{q^2}{(l+1)^2}\,,
\label{pot_Coul_2}
\end{equation}
where again $x\in(0,\infty)$ and  $l$, $q$ are real parameters.
The normalized eigenfunctions are given also by
(\ref{eigfunc_pot_Coul}), and as before there exist only
the normalizable eigenfunction for $k=0$ if $l\in(-3/2,-1)$, $q>0$, 
and in the range $l\in(-1,\infty)$, $q<0$, for all $k=0,\,1,\,2,\,\dots\,.$
However, the corresponding eigenvalues for the potentials (\ref{pot_Coul_2})
are now
\begin{equation}
E_k^{l,q}=\frac{q^2}{(l+1)^2}-\frac{q^2}{(k+l+1)^2}\,, \qquad
k=0,\,1,\,2,\,\dots
\label{spec_Coul_2}
\end{equation}

As in previous examples, we use two members of the family
(\ref{pot_Coul_2}) with different values of $l$.
Following Example~\ref{ex_oscil_inter}, we think of using $V_{l+1,q}(x)$ 
as the initial potential and $V_{l,q}(x)$ as the intermediate one 
with $-3/2<l<-1$. The eigenfunction of the initial potential has to be
square-integrable so we must set $q<0$. This means that
the intermediate potential $V_{l,q}(x)$ will have 
\emph{no} square-integrable 
eigenfunctions.
 
One simple way to overcome this difficulty is just to change the sign of $q$ in
the intermediate potential, which is what we will do in this Example. 
The interesting point, however, is that it is even possible to use a non 
normalizable eigenfunction of $V_{l,q}(x)$ as the intermediate one,
leading to physically interesting results. We will see this in the next Example. 
{}From the analysis of these two examples it can be shown
that the range $l\in(-3/2,-1)$ is indeed the only possibility if 
we restrict $q$ to take the same absolute value in the initial and intermediate
potentials. 

Now, assuming that $q<0$, we calculate the difference 
$$
V_{l,-q}(x)-V_{l+1,q}(x)
=\frac{q^2}{(l+1)^2}-\frac{q^2}{(l+2)^2}-\frac{2(l+1)}{x^2}-\frac{4 q}{x}\,.
$$
The first two terms coincide with $E_1^{l,q}>0$. The third and fourth are
always positive for $x\in(0,\infty)$ if $l<-1$ and $q<0$. An appropriate
$\g(x)$ is therefore
$$
\g_{l,q}(x)
=\frac{x}{\sqrt{\frac{(2l+3) q^2 x^2}{(l+1)^2 (l+2)^2}-4 q x-2(l+1)}}\,.
$$
The spectra (\ref{spec_Coul_2}) of two members of the family
(\ref{pot_Coul_2}) with values of $l$ differing by one coincide only for
the ground state energy.  As $E_0^{l,q}=0$ for all $l,q$, we will
transform the ground state of $V_{l+1,q}(x)$ by using the ground state of
$V_{l,-q}(x)$, both of them with zero energy. 
The final potential is, after some calculations,
\ba
&&V^{im}_{l,q}(x)=V_{l+1,q}(x)-2\bigg(\frac{\g_{l,q}^{\prime}}{\g_{l,q}}v_0^{l,-q}
+\frac {d v_0^{l,-q}}{dx}\bigg)+\frac{\g_{l,q}^{\prime\prime}}{\g_{l,q}}\nonumber\\
&&\quad=\frac{2 q}{x}+\frac{(l+1)(l+2)}{x^2}+\frac{q^2}{(l+2)^2}        \nonumber\\
& &\quad\quad+\frac{2(l+1) q \{2(l+1)(l+2)^3+(2 l^2+6l+5) qx\}}
{2(l+1)^2 (l+2)^2(l+1+2qx) x-(2l+3) q^2 x^3}                    \nonumber\\
& &\quad\quad+\frac{4(l+1)^2 (l+2)^2 (2l+3) q^3 x^2}
{x \{2(l+1)^2 (l+2)^2(l+1+2qx) -(2l+3) q^2 x^2\}^2}             \nonumber\\
& &\quad\quad-\frac{2(l+1)^3(l+2)^2 q\{(2 l^3+10 l^2+10 l-1) q x
+ 4(l+1)^2(l+2)^2\}}
{x \{2(l+1)^2 (l+2)^2(l+1+2qx) -(2l+3) q^2 x^2\}^2}\,,          \nonumber
\ea
where $v_0^{l,-q}(x)=(1/\z_0^{l,-q}(x))d \z_0^{l,-q}(x)/dx$, as usual. 
The known eigenfunction, with zero energy, of the previous potential is
\ba
\eta_0^{l,q}(x)&=&
\g_{l,q}(x)\bigg(-\frac{d\z_0^{l+1,q}(x)}{dx}+v_0^{l,-q}\z_0^{l+1,q}(x)\bigg)
\nonumber\\
&=&-\frac{2^{l+1}|q|^{l+5/2} 
e^{\frac{qx}{l+2}} x^{l+2}\{(l+1)(l+2)+(2l+3) q x\}}
{(l+1)(l+2)^{l+4}\sqrt{\Gamma(2l+4)}
\sqrt{\frac{(2l+3) q^2 x^2}{(l+1)^2 (l+2)^2}-4 q x-2(l+1)}}\,.
\nonumber
\ea
Since $l\in(-3/2,-1)$ and $q<0$, this function has neither zeros nor
singularities in $(0,\infty)$. Moreover it is square-integrable, for
the integral
$$
(\eta_0^{l,q},\eta_0^{l,q})=\int_0^\infty |\eta_0^{l,q}(x)|^2\,dx\,,
$$  
becomes after the change of variable $t=2|q|x/(l+2)$, 
$$
\frac{1}{2 (l+2)\Gamma(2l+4)}
\{4(l+1)I_1(l)-4(l+1)(2l+3)I_2(l)+(2l+3)^2 I_3(l)\}\,,
$$
where
$$
I_k(l)=\int_0^\infty\frac{e^{-t}t^{2l+3+k}}{d(l,t)}\,dt\,,
\quad\quad k=1,\,2,\,3,
$$
and $d(l,t)=(3+2l) t^2+8(l+2)(l+1)^2 t-8(l+1)^3$.  These integrals
converge when $l\in(-3/2,-1)$. We have computed numerically the complete
expression and checked that it takes positive real values in the same
interval. The result is a function strictly increasing with $l$, varying
from approximately 0.4 to 1. Taking into account these properties, the
eigenfunction $\eta_0^{l,q}(x)$ should be the ground-state of the image
potential.  
\end{example}

\begin{example}
\label{ex_Coul_2}
As our final example we will consider the previous one but using 
a non square-integrable eigenfunction, but without zeros,
of the intermediate potential.
As sometimes happens for the standard intertwining technique, we will
arrive to a physically meaningful image potential (see, for example, 
\cite{FerHusMiel98,FerHus99}).

Consider again the family of potentials (\ref{pot_Coul_2}). 
We choose $V_{l+1,q}(x)$ as the original potential, 
with $l\in(-3/2,-1)$ and $q<0$. The potential 
$V_{l,q}(x)$ will be the intermediate one. Their associated spectra
coincide just at zero energy, although the corresponding eigenfunction for
the intermediate potential is not square-integrable.
If we consider the difference
$$
V_{l,q}(x)-V_{l+1,q}(x)
=\frac{q^2}{(l+1)^2}-\frac{q^2}{(l+2)^2}-\frac{2(l+1)}{x^2}\,,
$$
we see again that the first two terms coincide with $E_1^{l,q}>0$ and
that the third one is always positive for $x\in(0,\infty)$ if $l<-1$, so
we can define
$$
\g_{l,q}(x)
=\frac{x}{\sqrt{\frac{(2l+3) q^2 x^2}{(l+1)^2 (l+2)^2}-2(l+1)}}\,.
$$

Now, we transform the ground state of $V_{l+1,q}(x)$ by using the formal
mathematical eigenfunction of $V_{l,q}(x)$ with zero eigenvalue, which is
not normalizable and has no zeros. The final potential becomes now
\ba
&&V^{im}_{l,q}(x)=V_{l+1,q}(x)
-2\bigg(\frac{\g_{l,q}^{\prime}}{\g_{l,q}}v_0^{l,q}+
\frac {d v_0^{l,q}}{dx}\bigg)
+\frac{\g_{l,q}^{\prime\prime}}{\g_{l,q}}
=\frac{q^2}{(l+2)^2}                            \nonumber\\
& &\quad\quad+\frac{2 q}{x}+\frac{(l+1)(l+2)}{x^2}                      
-\frac{2(l+1)q\{2(l+1)(l+2)^2+(2l+3)q x\}}{2(l+1)^3(l+2)^2 x-(2l+3)q^2 x^3}
\nonumber\\
& &\quad\quad+\frac{6(l+1)^3(l+2)^2(2l+3)q^2}
{\{2(l+1)^3(l+2)^2-(2l+3) q^2 x^2\}^2}\,,       \nonumber
\ea
where $v_0^{l,q}(x)=(1/\z_0^{l,q}(x))\,d\z_0^{l,q}(x)/dx$.  The known
eigenfunction with zero energy for the image potential is of the form
\ba
\eta_0^{l,q}(x)&=&
\g_{l,q}(x)\bigg(-\frac{d\z_0^{l+1,q}(x)}{dx}
+v_0^{l,q}\z_0^{l+1,q}(x)\bigg)                 \nonumber\\
&=&-\frac{2^{l+1}|q|^{l+5/2} 
e^{\frac{qx}{l+2}} x^{l+2}\{(l+1)(l+2)-q x\}}
{(l+1)(l+2)^{l+4}\sqrt{\Gamma(2l+4)}
\sqrt{\frac{(2l+3) q^2 x^2}{(l+1)^2 (l+2)^2}-2(l+1)}}\,.        \nonumber
\ea
As $l\in(-3/2,-1)$ and $q<0$, $\eta_0^{l,q}(x)$ has no singularities for
$x\in(0,\infty)$ but \emph{has} a zero at the value
$x_0=(l+1)(l+2)/q>0$. This function is square-integrable, since the
integral
$$
(\eta_0^{l,q},\eta_0^{l,q})=\int_0^\infty |\eta_0^{l,q}(x)|^2\,dx\,
$$  
becomes after the change of variable $t=2|q|x/(l+2)$ 
$$
\frac{1}{2 (l+2)\Gamma(2l+4)}
\{4(l+1)^2 I_1(l)+4(l+1)I_2(l)+I_3(l)\}\,,
$$
where 
$$
I_k(l)=\int_0^\infty\frac{e^{-t}t^{2l+3+k}}{d(l,t)}\,dt\,,
\quad\quad k=1,\,2,\,3,
$$
and now $d(l,t)=(3+2l) t^2-8(l+1)^3$.
These integrals can be computed explicitly with the aid of 
\cite[Formula 8.389.6]{GraRyz}:
\ba
&&\int_0^\infty\frac{t^\nu e^{-\mu t}}{\beta^2+t^2}\,dt
=\frac{\Gamma(\nu)}{2}\beta^{\nu-1}
\{e^{i(\mu\beta+(\nu-1)\pi/2)}\Gamma(1-\nu,i\beta\mu)           \nonumber\\
& &\quad\quad\quad\quad
+e^{-i(\mu\beta+(\nu-1)\pi/2)}\Gamma(1-\nu,-i\beta\mu)\}\,,     \nonumber\\
& &\quad\quad\quad\quad\quad\quad\quad\quad\quad\quad\quad\quad\quad\quad 
\mbox{Re}\,\beta>0\,,\ \mbox{Re}\,\mu>0\,,\ \mbox{Re}\,\nu>-1\,.\nonumber
\ea
In our case, $\mu=1>0$, $\beta=\sqrt{-8(l+1)^3/(2l+3)}$ is real and
positive for $l\in(-3/2,-1)$ and $\nu$ is alternatively $2l+4$, $2l+5$
and $2l+6$, all of them greater than $-1$. The final expression for
$\int_0^\infty |\eta_0^{l,q}(x)|^2\,dx$ is
$$
\frac{4(l+1)^2 i_1(l)+8(l+1)(l+2)i_2(l)
+2(2l+5)(l+2)i_3(l)}{4 (l+2)(2l+3)}\,,                  
$$
where
\ba
&&i_k(l)=\b(l)^{2l+2+k}
\{e^{i g(l,k)}\Gamma(-2l-2-k,i\b(l))                    \nonumber\\
&&\quad\quad+e^{-i g(l,k)}\Gamma(-2l-2-k,-i\b(l))\}\,,
\quad\quad\quad\quad k=1,\,2,\,3\,,                     \nonumber
\ea
being $g(l,k)=\b(l)+(2 l+2+k)\pi/2$ and $\b(l)=\sqrt{-8(l+1)^3/(2l+3)}$. 
This function is real, positive and strictly decreasing from
approximately 3 to 1 with $l\in(-3/2,-1)$. Then, the calculated
eigenstate should correspond to the first excited state of the final potential.
This implies that there should exist a ground state eigenfunction with
negative energy eigenvalue.  
\end{example}

\section{Conclusions and outlook}

Along this article we have studied several important facts concerning
the problem of $A$--related or intertwined Hamiltonians. To this end, we
have used simple but powerful group theoretical ideas.

The most important results are the the following. Firstly, we have
established the relationship between the finite difference algorithm
used in \cite{FerHusMiel98} and the transformation group on the set
of Riccati equations considered in \cite{CarRam99}, and we have shown that
the first is a particular instance of the second. 

Secondly, we have identified the group elements preserving a subset of
Riccati equations obtained from the set of Schr\"odinger-like equations by
means of the reduction procedure explained in Section~3. In this way we
have generalized the results of the finite difference algorithm to a
situation which seems to be new.

Thirdly, we have approached the problem of $A$--related or intertwined
Hamiltonians in terms of the transformation group on the set of Riccati
equations and the reduction method of Section~3, giving new insight into
the nature of the problem. 

Finally, we have illustrated by means of some examples the use of
the new general Theorems found at Section 5, thus generating potentials for
which one eigenfunction and its corresponding eigenvalue are exactly
known. As far as we know, some of these potentials have not 
been considered in the literature until now.

These results induce new interesting questions. For instance, in 
the case of not constant $\g(x)$,  
we do not know whether it is possible 
to factorize somehow the initial, intermediate and final
Hamiltonians. 
 
Another point concerns new applications of Theorem~\ref{my_theor2}. On the
one hand, it is an obvious idea to try to use potentials different from the
oscillator-like and Coulomb-like potentials treated here. 
On the other hand, inspired
by the Example~\ref{ex_Coul_2}, it is natural to consider explicitly other
non square-integrable eigenfunctions, without zeros, of the intermediate
potential, and see whether they lead
to physically meaningful final potentials. This will
involve to find the general solution of the intermediate
Schr\"odinger-like equation and to restrict then the discussion to the
particular solutions without zeros in the interval of interest, even if they
have no physical interpretation. This means, in some sense, 
to adapt to our current method the idea introduced by Mielnik in 
\cite{Mie}, and developed later in \cite{Fer84,Nie84,Don87,DiaNegNieRos99}, 
among other articles.

Finally, a finite difference formula has been used by Adler 
in order to discuss the B\"acklund transformations of the Painlev\'e
equations \cite{Adl93,Adl94}, also related with what are called 
{\em dressing chains} and the well-known Korteweg--de Vries equation, see
\cite{Sha92,VesSha93} and references therein. Moreover, the Darboux
transformation can be generalized by using more than one intermediate
eigenfunction of the original problem \cite{Cru55}. Likewise, there
exist generalizations of the usual intertwining technique 
to spaces with dimension greater 
than one \cite{AndBorIof84,AndBorIof84b,AndBorIofEid84}.
The natural question is whether there is some
relationship between these topics and the transformation group on
the set of Riccati equations discussed in this paper. 

We hope to provide some answers to these questions in future articles.

\begin{acknowledgments}

We thank the referee for valuable suggestions and
for calling several additional references to our attention.
We are also grateful to M. Asorey and B. Mielnik for useful 
comments and discussions.

A.R. thanks the Spanish Ministerio de Educaci\'on y Cultura
for a FPI grant, research project PB96--0717. Support of the
Spanish DGES (PB96--0717) and CONACYT project 32086-E (M\'exico), 
is also acknowledged. 

\end{acknowledgments}

%% Not optional, necessary:
\vfill\eject

%% This command is necessary! ==>>
\end{article}
\end{document}